\documentclass[twocolumn,pre,epfs,aps,showpacs]{revtex4}
\usepackage{epsfig}
\begin{document}
\title{Memory effects on the statistics of fragmentation}

\author{L. E. Araripe, J. S. Andrade Jr., and R. N. Costa Filho}
\affiliation{$^1$Departamento de F\'{\i}sica, Universidade Federal do
Cear\'a, 60451-970 Fortaleza, Cear\'a, Brazil.}
\date{\today}

\begin{abstract}
We investigate through extensive molecular dynamics simulations the
fragmentation process of two-dimensional Lennard-Jones systems. After
thermalization, the fragmentation is initiated by a sudden increment
to the radial component of the particles' velocities. We study the
effect of temperature of the thermalized system as well as the
influence of the impact energy of the ``explosion'' event on the
statistics of mass fragments. Our results indicate that the cumulative
distribution of fragments follows the scaling ansatz $F(m)\propto
m^{-\alpha}\exp{[-(m/m_0)^\gamma]}$, where $m$ is the mass, $m_0$ and
$\gamma$ are cutoff parameters, and $\alpha$ is a scaling exponent
that is dependent on the temperature. More precisely, we show clear
evidence that there is a characteristic scaling exponent $\alpha$ for
each macroscopic phase of the thermalized system, i.e., that the
non-universal behavior of the fragmentation process is dictated by
the state of the system before it breaks down.
\end{abstract}

\maketitle

\section{Introduction}
The process of breaking solids into smaller pieces has been the
subject of deep thoughts since the time of the Greeks, who tried to
understand the building blocks of matter. Not going so far away in
time or neither in the area of particle physics, the fragmentation
process is still an important problem to study since it is a main
issue in current problems in our day-to-day life. For instance, to
understand why or how a material breaks is relevant in the development
of new technological devices or in geological problems
\cite{Turcotte86,Wittel04}. Because it is such a significant issue,
a large number of experiments in fragmentation have been performed in
order to collect data of fractures in many types of materials and
 objects
forms
 \cite{Turcotte86,Wittel04,Oddershede93,Meibom96,Ishii92,Kadono97}.
The number of theoretical articles on this topic is no smaller. The
main focus of recent studies in this field is based on molecular
dynamics (MD) simulations, where the results show an ubiquitous
 scaling
behavior in the distribution of the mass fragments, $F(m)\sim
m^{-\alpha}$, with the exponent $\alpha$ depending on the
dimensionality and initial parameters of the system
\cite{Oddershede93,Meibom96,Marsili96}.

The aforementioned experimental and theoretical studies have shown
that the mass distribution belongs to the same universality class for
large enough input energies when the MD system breaks into smaller
pieces \cite{Turcotte86,Oddershede93,Kadono97,Inaoka97,Katsuragi03}. 
However, using a molecular dynamics approach Ching et al.
 \cite{Ching99}
fragmented an object represented as a set of particles interacting via
 
the Lennard-Jones potential with the fracture process being induced by
 
random initial velocities assigned to the particles. The resulting 
steady-state form found for the cumulative mass distribution displays 
a typical power-law region, with a non-universal exponent that
 increases
with the total initial energy given to the system. The same behavior
has been observed in experimental fragmentation of long glass rods 
\cite{Ching00} and duly interpreted as an indication that the 
fragmentation process is not a self-organizing phenomenon, contrary 
to the assumption of Oddershede et al. \cite{Oddershede93}
s
In contrast to the self-organized criticality paradigm where the power
law behavior should appear without a control parameter, there is a
interesting claim that, in impact fragmentation, criticality could be
tuned at a nonzero impact energy \cite{Herrmann99}. In this way, the
fragment-size distribution should satisfy a scaling form similar to
that of the cluster-size distribution of percolation clusters
\cite{Stauffer}, but belonging to another universality class 
\cite{Astrom00,Katsuragi03}. From the results of such numerical models
 
it has been suggested that there exists a critical imparted energy,
below which the object to be fragmented is only damaged, and above
which it breaks down into numerous smaller pieces. The transition
between the \emph{damaged} to the \emph{fragmented} states behaves as
a critical point, with the fragment size distribution displaying a
scaling form similar to that described in percolation theory
\cite{Stauffer}. The same dependence on the impact energy for the
fragmented state has been found very recently in another numerical 
model for the fragmentation of a circular disk by projectiles
\cite{Behera04} as well as in the experimental fragmentation of shells
\cite{Wittel04}.

In the present work our aim is to investigate through molecular
dynamics simulations the effect of different initial conditions (e.g.,
temperature and impact energy) on the mass distribution of fragments
generated after an ``explosion'' takes place. One of our goals is to
show that the scaling behavior observed in the statistics of mass
fragments is non-universal and that this non-universality has a direct
correspondence with the state of the system prior to fragmentation
process. In Section II we describe the details of the model and
simulations. The results are shown in Section III, while the
conclusions and some perspectives are presented in Section IV.

\section{Model}

The fragmentation model used here is based on the one described in
Refs.~\cite{Naftaly,Diehl00}. The initial state of the object to be
fragmented is a thermalized configuration generated through a standard
molecular dynamics simulation in the microcanonical ensemble. The
particles interact with each other through a 6-12 Lennard-Jones pair
potential and the system is brought to the desired equilibrium
temperature by integration of Newton's equations of motion
\cite{Rapaport}. A neighbor-list method is applied and periodic
boundary conditions are used in all directions. This allows us to
simulate up to $10^5$ particles for a single realization of the
fragmentation system. The results are then taken from an average of
fifty realizations (the direction of the initial velocities for the
particles are different for each sample) for a given set of initial
conditions, as defined by the value of the temperature, particle
density, and energy given to break the MD system apart. This
 ``explosion
energy'' is specified through the parameter $R$, defined as the ratio
between the initial kinetic energy and the initial potential energy of
the particle, immediately after the velocities are set according to
the equation below
\begin{equation}
{\mathbf{v}}_i(0)={\mathbf{v}}_i^T+C{\mathbf{r}}_i(0),
\label{pertubav}
\end{equation}
where ${\mathbf{v}_i^T}$ are the initial velocities and
${\mathbf{r}}_i(0)$ are the initial positions of the particles,
obtained in the thermalization stage. The second term in the above
equation is responsible for an expansion process that is preceded by
an explosive event. The proportionality constant $C$ has units of
inverse of time and gives a measure of the initial energy imparted to
the object. From time zero onward, no energy is added to the system
and the particles' positions and velocities are now calculated
considering free boundary conditions. As a result, the system expands
and the particles are distributed among clusters (fragments) of
different masses. Each particle is considered as a monomeric cluster
with unitary mass. There are several definitions for a particle
cluster \cite{Sator}. Here, two particles will belong to the same
cluster if they are separated by a distance smaller than an arbitrary
cutoff, $r_c=3\sigma$. The fragments are then classified according to
their mass $m$ and counted to enable the calculation of the
distributions $n(m)$ and $F(m)$, both normalized here by the total
number of fragments.

\begin{figure}[h]
\includegraphics[width=7.5cm]{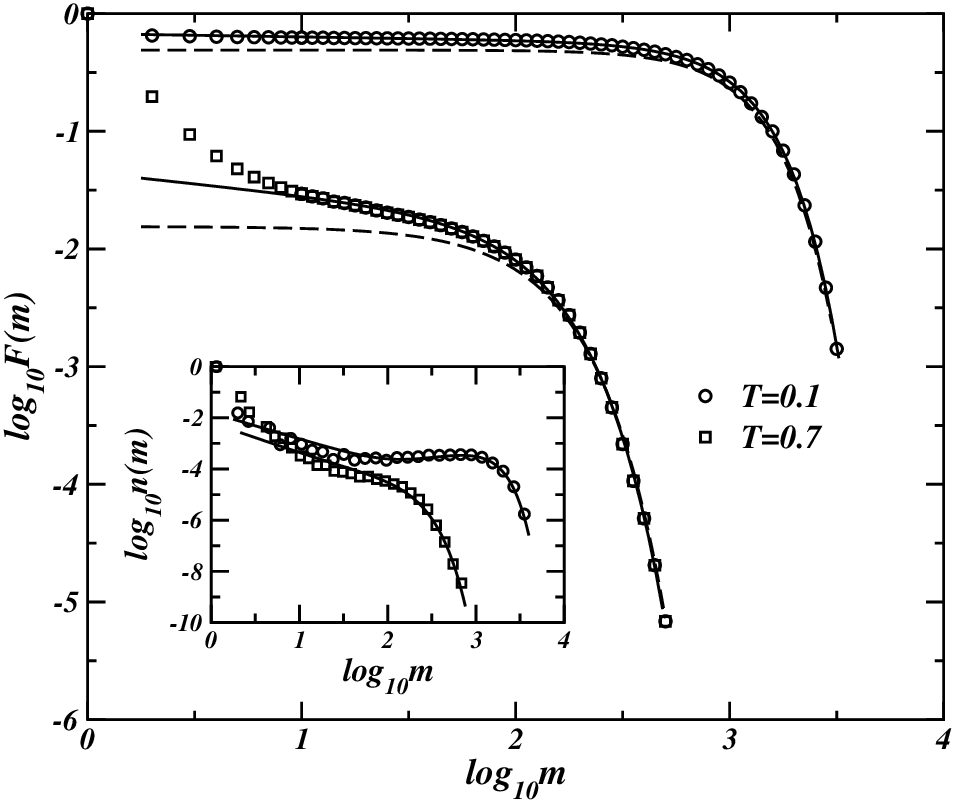}
\caption{The cumulative mass distribution of fragments $F(m)$ for
$\rho=0.61$, $R=0.43$, and two different values of temperature
$T=0.1$ (circles) and $0.7$ (squares). For comparison the dashed line
corresponds to $f(m)\sim\exp{[-(m/m_0)^\gamma]}$. The inset shows the
corresponding mass distribution $n(m)$, with the solid line
representing the best fit.} \label{fig1} \end{figure}

\section{Results}

Due to the fluctuations in $n(m)$, it is usually more convenient to
work with the cumulative form of the mass distribution defined as
\cite{Oddershede93}

\begin{equation}
F(m)=\int_m^\infty n(m')dm'.
\end{equation}
In Fig.~\ref{fig1} we show the behavior of $F(m)$ for fixed values of
the system density $\rho=0.61$ and the parameter $R=0.43$, and two
different values of the temperature $T$. As can be seen, both
distributions display a region of power-law behavior at intermediate
values of $m$ followed by a typical cutoff due to finite size.
From Ref.~\cite{Diehl02}, the following expression has been proposed
to describe the behavior of $F(m)$:

\begin{equation}
F(m)\sim m^{1-\alpha}\exp{[-(m/m_0)^\gamma]},
\label{fit}
\end{equation}
where $\alpha$ is a scaling exponent, and $m_0$ and $\gamma$ are
cutoff parameters. As depicted in Fig.~\ref{fig1} the application of a
standard non-linear estimation algorithm to both data sets shows that
Eq.~(\ref{fit}) fits well the scaling region for intermediate masses
as well as the decaying cutoff for large fragment sizes, which is
compatible with a stretched exponential behavior. This fitting
procedure gives estimates for the scaling exponent that are
substantially different, namely, $\alpha=1.02\pm 0.01$ and $\alpha=
 1.20\pm 0.01$
for $T=0.1$ and $0.7$, respectively. There is however a discrepancy
between the data and the curve fitted with Eq.~(\ref{fit}) at $T=0.7$
for the region of small fragments. This can be readily explained in
terms of the large ``evaporation'' rates at high values of the
temperature -- an expected effect that is responsible for the
progressive detachment of small clusters from the hull of large and
medium fragments after the explosion event. The inset of
Fig.~\ref{fig1} shows that the corresponding behavior of the
distribution $n(m)\equiv dF(m)/dm$ for both values of $T$ is also
consistent with scaling ansatz Eq.~(\ref{fit}), when the parameter
used are the same as those obtained for fitting its integral form
$F(m)$.

\begin{figure}
\includegraphics[width=7.5cm]{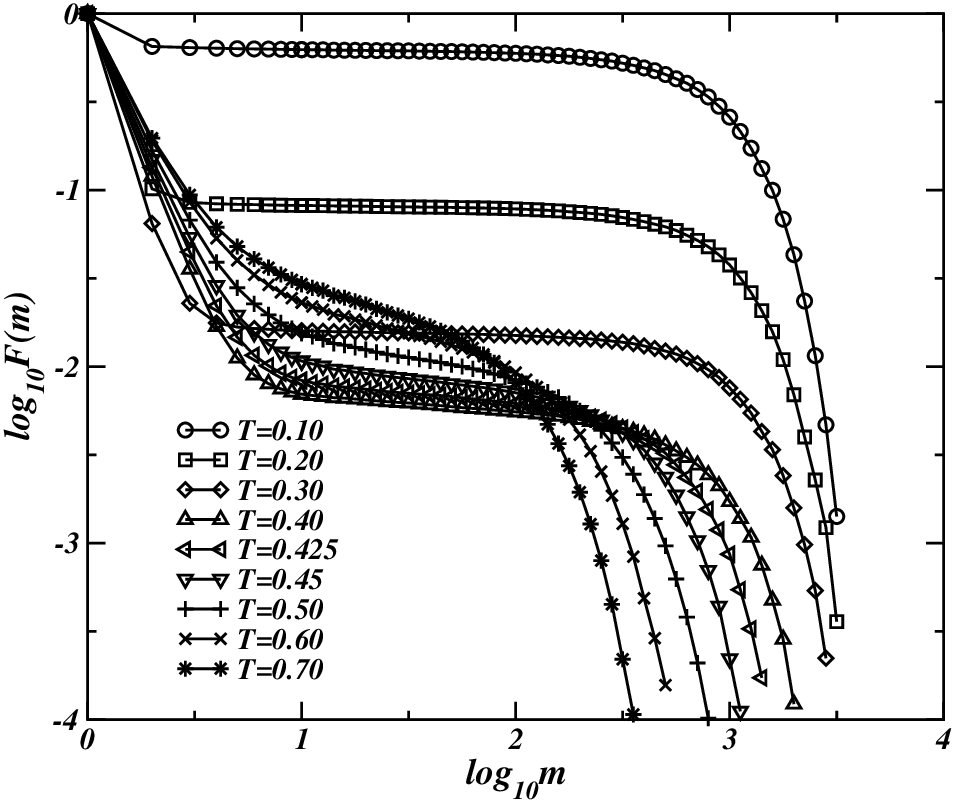}
\caption{Log-log plot of $F(m)$ for $\rho=0.61$, $R=0.43$,
and different values of temperature.} \label{fig2} \end{figure}

\begin{figure}
\includegraphics[width=7.5cm]{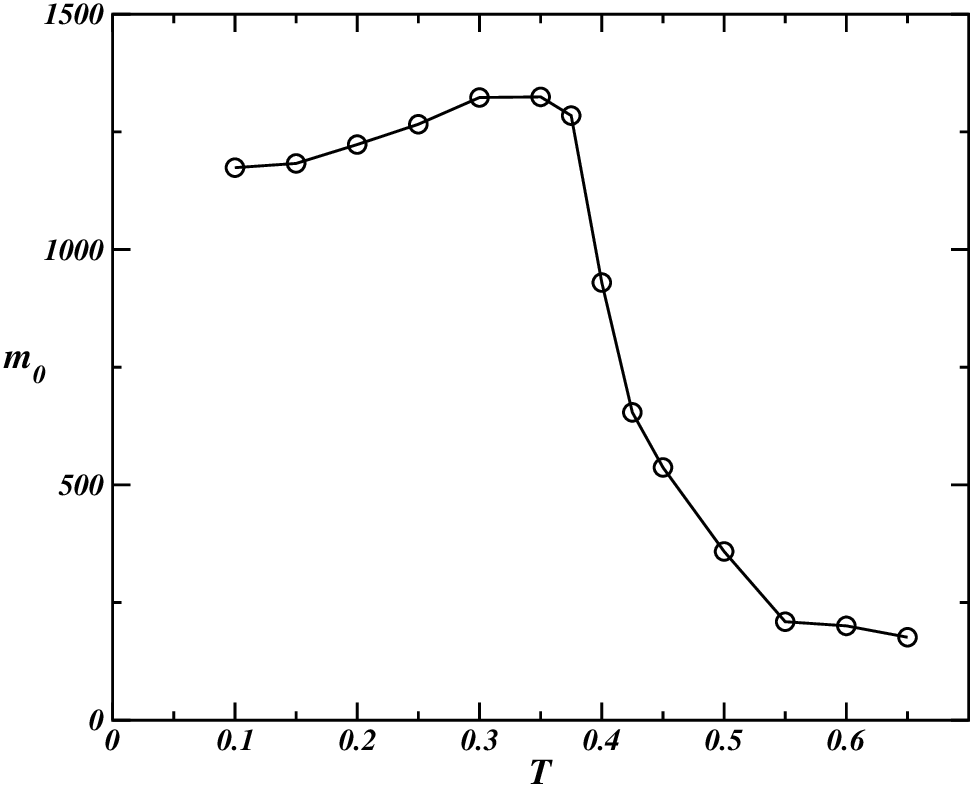}
\caption{The behavior of the parameter $m_0$ against temperature. The
values of density and the parameter $R$ are the same as in the
previous figure.}\label{fig3}  \end{figure}

In Fig.~\ref{fig2} we show the profiles of the distribution $F(m)$ for
several values of the temperature in the range $0.1 \le T \le 0.7$ and
a fixed value of $\rho=0.61$. When observed in detail, the diversity
in shape of $F(m)$ for intermediate and large fragment sizes indicate
that the fragmentation process must be restricted to a discrete and
much smaller number of different classes of behavior than the
variation with an entire spectrum of thermalization temperatures could
suggest. This fact is quantitatively verified when we observe that,
after fitting Eq.~(\ref{fit}) to each data set, the scaling parameter
$\alpha$ can assume one among only three distinct numerical values for
the distributions generated at nine different temperatures.  In
Fig.~\ref{fig3} we show the variation of the cutoff parameter $m_0$
with temperature for $\rho=0.61$. As the temperature increases from
$T=0.1$, $m_0$ remains approximately constant up to $T \approx 0.375$,
where it suddenly drops to become again constant, at least up to the
maximum value of the temperature we use in our simulations, $T=0.7$.
This sharp transition in $m_0$ indicates the existence of a
``critical'' temperature below which a large cluster (i.e., with a
size of the order of the system size) can exist.

\begin{figure}
\includegraphics[width=7.5cm]{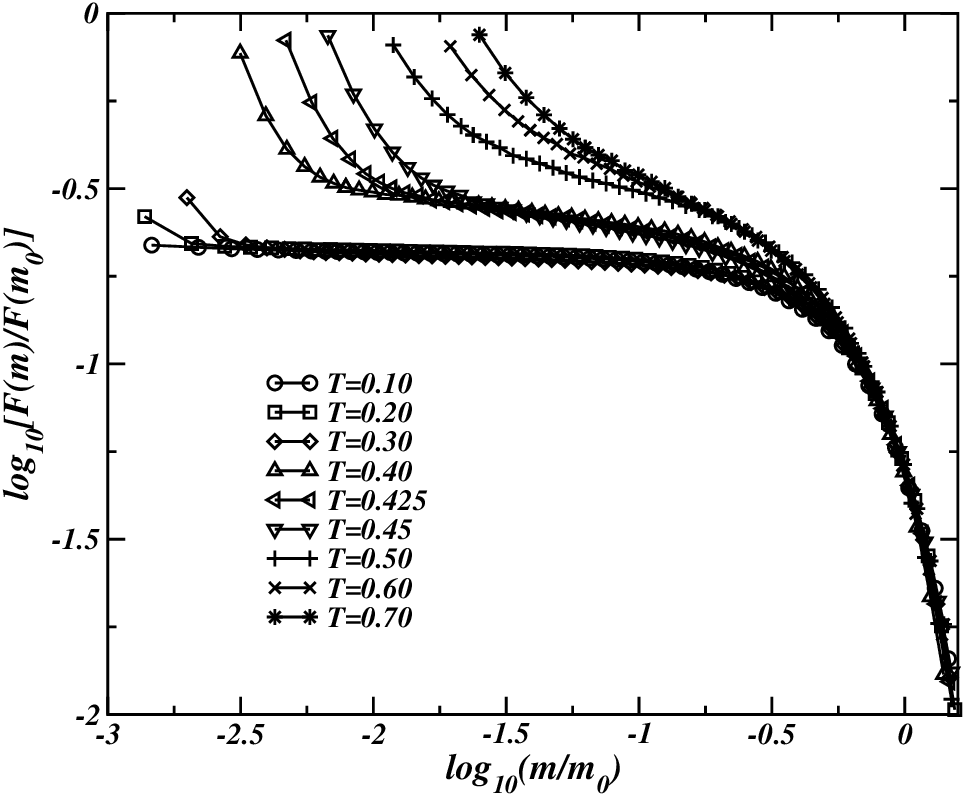}
\caption{Data collapse of the distribution $F(m)$ shown in
Fig.\ref{fig2}. The collapse has been obtained by rescaling the
abscissas $m$ of each distribution to its corresponding cutoff
parameters $m_0$, as well as rescaling the values $F(m)$ to $F(m_0)$.}
\label{fig4} \end{figure}

\begin{figure}
\includegraphics[width=7.5cm]{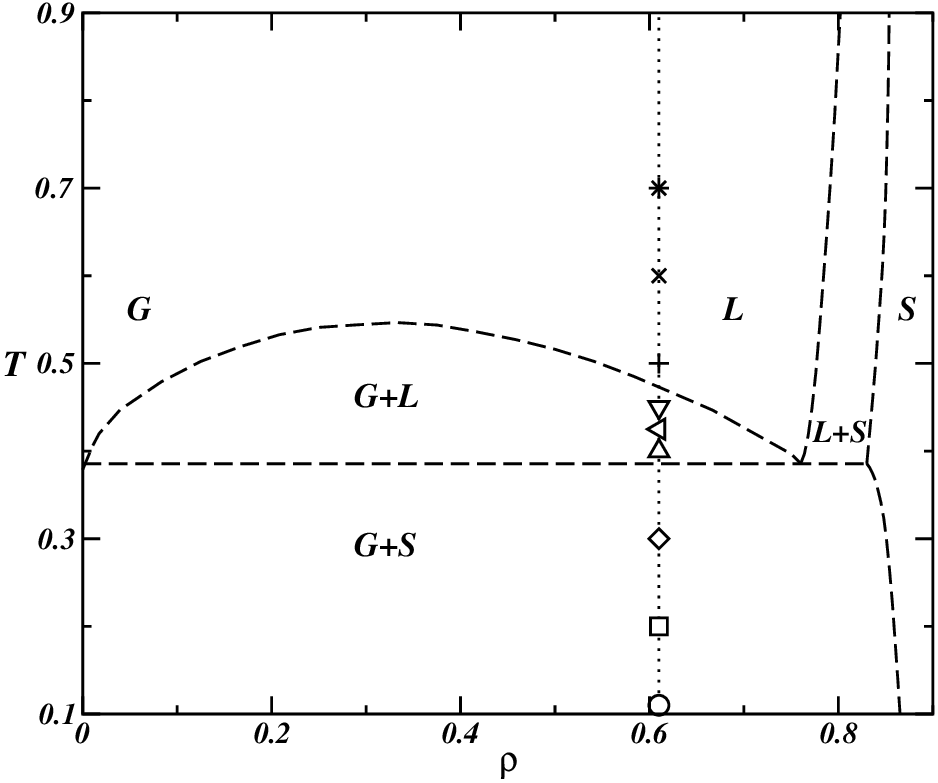}
\caption{The phase diagram for a two-dimensional system with particles
interacting through  the Lennard-Jones potential. The points following
the vertical dashed line represent the values of temperature used in
our simulations.}  \label{fig5} \end{figure}

In Fig.~\ref{fig4} we show the data collapse obtained by rescaling the
abscissas $m$ of each curve shown in Fig.~\ref{fig3} to its
corresponding estimate of the cutoff parameter $m_0$, as well as
rescaling the values $F(m)$ to $F(m_0)$. This results clearly reveal
the presence of only three groups of distributions, each one with a
characteristic value for the scaling exponent $\alpha$. Such a
 behavior
can be explained with the help of the phase diagram shown in
Fig.~\ref{fig5}, where the points following the vertical dashed line
represent the values of temperature used in our simulations. From this
diagram, we readily deduce that the threefold statistics of mass
fragments shown in Fig.~\ref{fig4} is a direct consequence of the
three distinct phases to which the thermalized objects belonged before
they have been broken apart. It is interesting to note that, although
the collapses are rather convincing for intermediate and large
 fragment
sizes, the apparent divergence characterizing the statistics of small
fragments due to ``evaporation'' appears to be continuously changing
with temperature within each of the three groups of collapsed data.

\begin{figure}
\includegraphics[width=7.5cm]{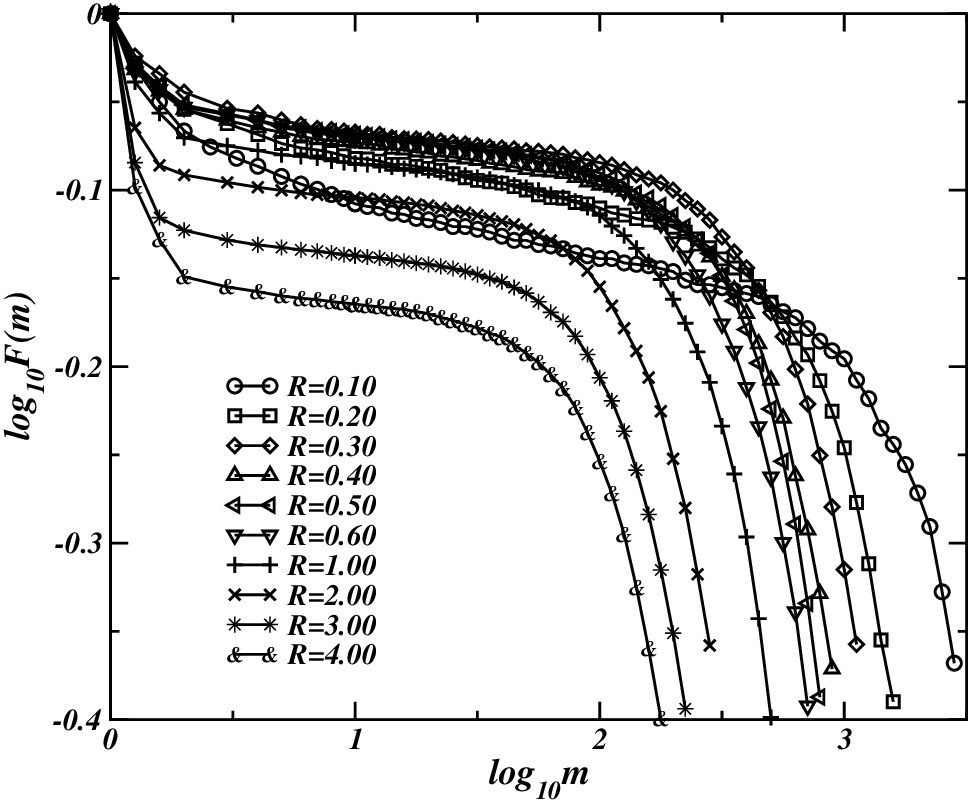}
\caption{Log-log plot of the cumulative distribution $F(m)$ for values
of the energy parameter $R$ ranging from $R=0.1$ to $4.0$. The MD
systems have been thermalized with $T=0.1$ and $\rho=0.61$}
\label{fig6} \end{figure}

\begin{figure}
\includegraphics[width=7.5cm]{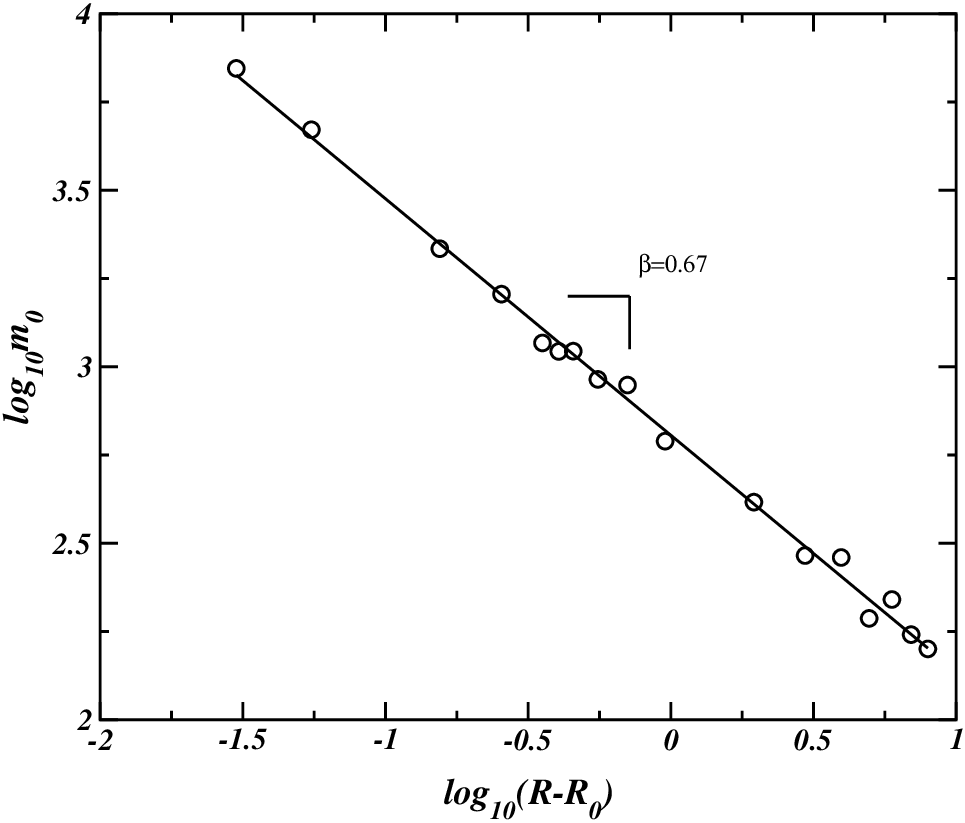}
\caption{Log-log plot of the crossover parameter $m_0$ against the
energy difference $(R-R_0)$ for MD systems thermalized with $\rho
=0.61$ and $T=0.1$.} \label{fig7} \end{figure}

\begin{figure}[h]
\includegraphics[width=7.5cm]{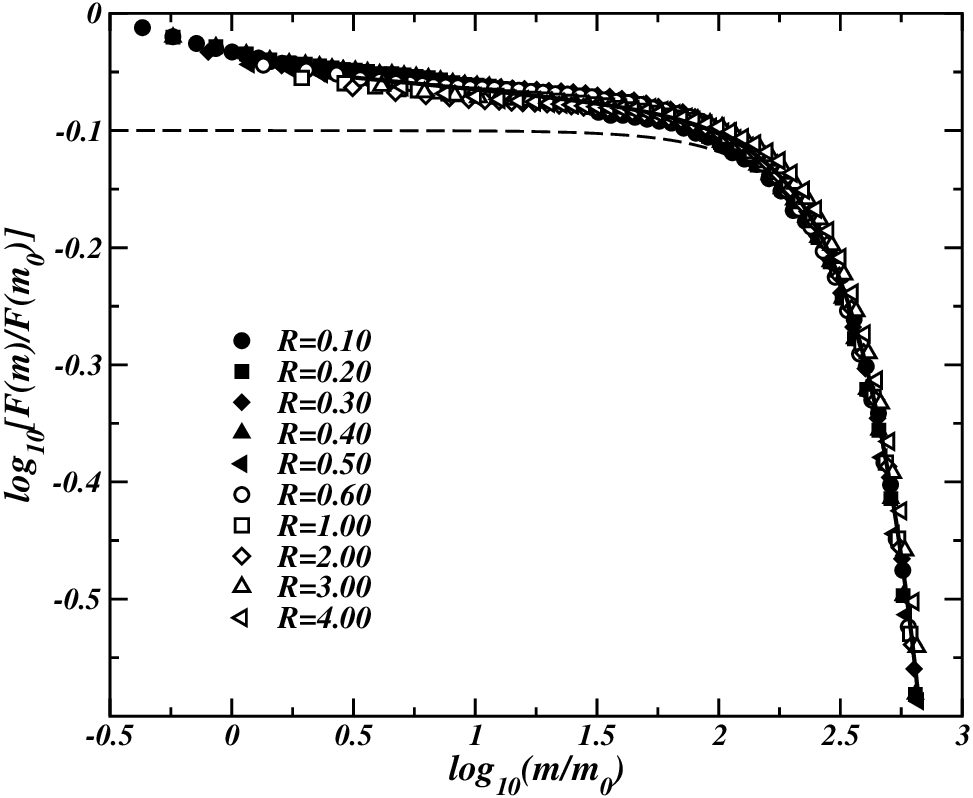}
\caption{Data collapse of the distribution $F(m)$ for $\rho=0.61$,
$T=0.1$, and different values of the energy input $R$. For comparison
the dashed line corresponds to $f(m)\sim\exp{[-(m/m_0)^\gamma]}$.}
\label{fig8}
\end{figure}

The situation becomes entirely different when we analyze the influence
of the energy parameter $R$ on the statistics of the fragmentation
process. In Fig.~\ref{fig6} we show the distributions $F(m)$ computed
for MD systems thermalized with temperature $T=0.1$, particle density
$\rho=0.61$, and for different values of $R=0.1$, $0.2$, $0.3$, $0.4$,
 $0.5$, $0.6$,
$1.0$, $2.0$, $3.0$, and $4.0$. From the non-linear fitting of
Eq.~\ref{fit} to each data set we notice that, while the scaling
exponent remains approximately constant at $\alpha \approx 1.02$,
$m_0$ changes significantly with $R$. Precisely, as shown in
Fig.~\ref{fig7}, the decay of $m_0$ with $R$ can be described in terms
of a power-law \begin{equation} m_0=a(R-R_0)^{-\beta}, \label{m0}
\end{equation}
where $a=640.0\pm 0.1$ is a prefactor and the exponent $\beta=0.67
\pm 0.02$. The parameter $R_0$ is an offset that is related to the
competition between the thermal energy of motion and the energy that
holds the system together, i.e., the ratio between the kinetic energy
and the potential energy just before the velocities are settled
according to Eq.~\ref{pertubav} and the boundary is lifted. Using
Eq.~\ref{m0} and its estimated parameters to rescale the data
presented in Fig.~\ref{fig6}, we show in Fig.~\ref{fig8} that the
distributions for all values of $R$ can be nicely represented by a
single data-collapsed curve. Of course, this should only be valid for
systems subjected to the same thermalization process, i.e., if $\rho$
and $T$ are kept constant.

\section{Conclusions}
In summary, we performed an extensive study of a two-dimensional
fragmentation process through molecular dynamics simulations.
Specifically, we have shown how the statistics of the fragmentation
process depends on ({\it i}) the thermalization temperature of the
system before its breakdown, and ({\it ii}) the energy imparted to the
system to induce fragmentation. In the first case, we verified that
the cumulative mass distribution follows a power law for intermediate
masses, with an exponent that depends on the region of temperature
considered. More precisely, we showed that it is the phase of the
thermalized object that is responsible for the difference in these
scaling exponents. It means that the process studied here can be
rather sensitive to the previous state of the system, although it
introduces a significant disturbance from an energetic point of view.
As a consequence, fragmentation carries memory. Finally, we turned our
attention to the variation of the parameter $R$. Differently to the
previous case, we obtained a unique scaling exponent for the
cumulative mass distribution for different values of $R$. This result
is in good agreement with previous studies in the literature
indicating some sort of universal behavior present in fragmentation
processes \cite{Behera04}.

\begin{acknowledgments}
This work has been supported by CNPq, CAPES and FUNCAP.
\end{acknowledgments}

\end{document}